\documentclass[12pt]{article}
\usepackage{graphics}
\usepackage{amssymb,epsfig,amsmath,euscript,array}

\newcounter{multieqs}

\newcommand{\be}{\begin{equation}}
\newcommand{\ee}{\end{equation}}
\newcommand{\eq}[1]{(\ref{#1})}

\newcommand{\ket}[1]{|#1 \rangle}

\newcommand{\bm}[1]{\mbox{\boldmath $#1$}}

\def\bd{\begin{document}}
\def\ed{\end{document}}
\def\nn{\nonumber}
\def\bea{\begin{eqnarray}}
\def\eea{\end{eqnarray}}
\let\bm=\bibitem
\let\la=\label

\def\npb#1#2#3{Nucl. Phys. {\bf{B#1}} #3 (#2)}
\def\plb#1#2#3{Phys. Lett. {\bf{#1B}} #3 (#2)}
\def\prl#1#2#3{Phys. Rev. Lett. {\bf{#1}} #3 (#2)}
\def\prd#1#2#3{Phys. Rev. {D \bf{#1}} #3 (#2)}
\def\cmp#1#2#3{Comm. Math. Phys. {\bf{#1}} #3 (#2)}
\def\cqg#1#2#3{Class. Quantum Grav. {\bf{#1}} #3 (#2)}
\def\nppsa#1#2#3{Nucl. Phys. B (Proc. Suppl.) {\bf{#1A}}#3 (#2)}
\def\ap#1#2#3{Ann. of Phys. {\bf{#1}} #3 (#2)}
\def\ijmp#1#2#3{Int. J. Mod. Phys. {\bf{A#1}} #3 (#2)}
\def\rmp#1#2#3{Rev. Mod. Phys. {\bf{#1}} #3 (#2)}
\def\mpla#1#2#3{Mod. Phys. Lett. {\bf A#1} #3 (#2)}
\def\jhep#1#2#3{J. High Energy Phys. {\bf #1} #3 (#2)}
\def\atmp#1#2#3{Adv. Theor. Math. Phys. {\bf #1} #3 (#2)}

\newcommand{\EQ}[1]{\begin{equation} #1 \end{equation}}
\newcommand{\AL}[1]{\begin{subequations}\begin{align} #1
\end{align}\end{subequations}}
\newcommand{\SP}[1]{\begin{equation}\begin{split} #1 \end{split}\end{equation}}
\newcommand{\ALAT}[2]{\begin{subequations}\begin{alignat}{#1} #2
\end{alignat}\end{subequations}}
\def\beqa{\begin{eqnarray}}
\def\eeqa{\end{eqnarray}}
\def\beq{\begin{equation}}
\def\eeq{\end{equation}}

\def\N{{\cal N}}
\def\sst{\scriptscriptstyle}
\def\thetabar{\bar\theta}
\def\Tr{{\rm Tr}}
\def\one{\mbox{1 \kern-.59em {\rm l}}}

\def\a{\alpha}      \def\da{{\dot\alpha}}
\def\b{\beta}       \def\db{{\dot\beta}}
\def\c{\gamma}  \def\C{\Gamma}  \def\cdt{\dot\gamma}
\def\d{\delta}  \def\D{\Delta}  \def\ddt{\dot\delta}
\def\e{\epsilon}        \def\vare{\varepsilon}
\def\f{\phi}    \def\F{\Phi}    \def\vvf{\f}
\def\h{\eta}
\def\k{\kappa}
\def\l{\lambda} \def\L{\Lambda}
\def\m{\mu} \def\n{\nu}
\def\o{\omega}
\def\p{\pi} \def\P{\Pi}
\def\r{\rho}
\def\s{\sigma}  \def\S{\Sigma}
\def\t{\tau}
\def\th{\theta} \def\Th{\Theta} \def\vth{\vartheta}
\def\X{\Xeta}
\def\z{\zeta}

\def\cA{{\cal A}} \def\cB{{\cal B}} \def\cC{{\cal C}}
\def\cD{{\cal D}} \def\cE{{\cal E}} \def\cF{{\cal F}}
\def\cG{{\cal G}} \def\cH{{\cal H}} \def\cI{{\cal I}}
\def\cJ{{\cal J}} \def\cK{{\cal K}} \def\cL{{\cal L}}
\def\cM{{\cal M}} \def\cN{{\cal N}} \def\cO{{\cal O}}
\def\cP{{\cal P}} \def\cQ{{\cal Q}} \def\cR{{\cal R}}
\def\cS{{\cal S}} \def\cT{{\cal T}} \def\cU{{\cal U}}
\def\cV{{\cal V}} \def\cW{{\cal W}} \def\cX{{\cal X}}
\def\cY{{\cal Y}} \def\cZ{{\cal Z}}

\def\ua{\underline{\alpha}}
\def\ub{\underline{\phantom{\alpha}}\!\!\!\beta}
\def\uc{\underline{\phantom{\alpha}}\!\!\!\gamma}
\def\um{\underline{\mu}}
\def\ud{\underline\delta}
\def\ue{\underline\epsilon}
\def\una{\underline a}\def\unA{\underline A}
\def\unb{\underline b}\def\unB{\underline B}
\def\unc{\underline c}\def\unC{\underline C}
\def\und{\underline d}\def\unD{\underline D}
\def\une{\underline e}\def\unE{\underline E}
\def\unf{\underline{\phantom{e}}\!\!\!\! f}\def\unF{\underline F}
\def\unm{\underline m}\def\unM{\underline M}
\def\unn{\underline n}\def\unN{\underline N}
\def\unp{\underline{\phantom{a}}\!\!\! p}\def\unP{\underline P}
\def\unq{\underline{\phantom{a}}\!\!\! q}
\def\unQ{\underline{\phantom{A}}\!\!\!\! Q}
\def\unH{\underline{H}}

\def\As {{A \hspace{-6.4pt} \slash}\;}
\def\bs {{b \hspace{-6.4pt} \slash}\;}
\def\Ds {{D \hspace{-6.4pt} \slash}\;}
\def\ds {{\del \hspace{-6.4pt} \slash}\;}
\def\ss {{\s \hspace{-6.4pt} \slash}\;}
\def\ks {{ k \hspace{-6.4pt} \slash}\;}
\def\ps {{p \hspace{-6.4pt} \slash}\;}
\def\pas {{{p_1} \hspace{-6.4pt} \slash}\;}
\def\pbs {{{p_2} \hspace{-6.4pt} \slash}\;}

\def\Fh{\hat{F}}
\def\Nh{\hat{N}}
\def\Vh{\hat{V}}
\def\Xh{\hat{X}}
\def\ah{\hat{a}}
\def\xh{\hat{x}}
\def\yh{\hat{y}}
\def\ph{\hat{p}}
\def\xih{\hat{\xi}}

\def\psit{\tilde{\psi}}
\def\Psit{\tilde{\Psi}}
\def\tht{\tilde{\th}}

\def\At{\tilde{A}}
\def\Qt{\tilde{Q}}
\def\Rt{\tilde{R}}
\def\Nt{\tilde{N}}

\def\at{\tilde{a}}
\def\st{\tilde{s}}
\def\ft{\tilde{f}}
\def\pt{\tilde{p}}
\def\qt{\tilde{q}}
\def\vt{\tilde{v}}
\def\nt{\tilde{n}}

\def\delb{\bar{\partial}}
\def\bz{\bar{z}}
\def\bD{\bar{D}}
\def\bB{\bar{B}}

\def\bk{{\bf k}}
\def\bl{{\bf l}}
\def\bp{{\bf p}}
\def\bq{{\bf q}}
\def\br{{\bf r}}
\def\bx{{\bf x}}
\def\by{{\bf y}}
\def\bR{{\bf R}}
\def\bV{{\bf V}}

\def\d{\delta}\def\D{\Delta}\def\ddt{\dot\delta}

\def\pa{\partial} \def\del{\partial}
\def\xx{\times}
\def\uno{\mbox{1 \kern-.59em {\rm l}}}

\def\trp{^{\top}}
\def\inv{^{-1}}
\def\dag{{^{\dagger}}}
\def\pr{^{\prime}}

\def\rar{\rightarrow}
\def\lar{\leftarrow}
\def\lrar{\leftrightarrow}

\newcommand{\0}{\,\!}      %this is just NOTHING!
\def\one{1\!\!1\,\,}
\def\im{\imath}
\def\jm{\jmath}

\newcommand{\tr}{\mbox{tr}}
\newcommand{\slsh}[1]{/ \!\!\!\! #1}

\def\vac{|0\rangle}
\def\lvac{\langle 0|}

\def\hlf{\frac{1}{2}}
\def\ove#1{\frac{1}{#1}}

\def\Box{\square}
\def\ZZ{\mathbb{Z}}
\def\CC#1{({\bf #1})}
\def\bcomment#1{}
\def\bfhat#1{{\bf \hat{#1}}}
\def\VEV#1{\left\langle #1\right\rangle}

\newcommand{\ex}[1]{{\rm e}^{#1}} \def\ii{{\rm i}}

\def\rr{{\rm r}} \def\rs{{\rm s}}\def\rv{{\rm v}}
\def\ri{{\rm i}}\def\rj{{\rm j}}

\newcommand{\lrbrk}[1]{\left(#1\right)}
\newcommand{\sfrac}[2]{{\textstyle\frac{#1}{#2}}}

\begin{document}

\hfill{hep-th/0301036}

\vspace{20pt}

\begin{center}

{\Large \bf Correspondence between the 3-point BMN \\}
\vspace{10pt}
{\Large \bf correlators and the 3-string vertex on the pp-wave}

\vspace{30pt}

{\bf Chong-Sun Chu$^{a,b}$ and Valentin V. Khoze$^{c}$}

{\small \em
\begin{itemize}
\item[$^a$] Centre for Particle Theory,
Department of Mathematical Sciences,\\
University of Durham, Durham, DH1 3LE, UK
\item[$^b$]Department of Physics, National Tsing Hua University,
Hsinchu, Taiwan 300, R.O.C.
\item[$^c$] Centre for Particle Theory,
Department of Physics and IPPP,\\
University of Durham, Durham, DH1 3LE, UK
\end{itemize}
}

\vspace{10pt}

Email: {\sffamily \tt chong-sun.chu, valya.khoze@durham.ac.uk }

\vspace{30pt} {\bf Abstract}

\end{center}

% ABSTRACT goes here

The PP-wave/SYM proposal in its original form emphasizes a duality relation 
between the masses of the string states and the anomalous dimensions of the
corresponding BMN operators in gauge theory, the 
{\it mass--dimension} type duality. 
In this paper, we give evidence in favour of another duality
relation of the {\it vertex--correlator} type,
which relates the coefficients of 3-point correlators of BMN operators
in gauge theory to 3-string vertices in lightcone string field theory in the
pp-wave background. We verify
that all the available field theory results in the literature, 
as well as the newly obtained ones, for the 3-point functions
are successfully reproduced from our proposal.

\vspace{0.5cm}

\setcounter{page}0
\thispagestyle{empty}
\newpage

%%%%%%%%%%%%%% ordinary document (end) ####################################

\section{Introduction}

In this paper, we continue the study of the correspondence 
\cite{Constable1,huang,CKT} between
3-point functions of BMN operators in $\cN=4$ Yang-Mills theory and 3-string
interactions in the pp-wave lightcone string field theory.

The pp-wave/SYM proposal in its original form \cite{BMN} 
emphasizes a duality relation
between the anomalous dimensions of the BMN operators and the masses of the
corresponding string states. Much progress had been made in verifying this relation 
in the planar limit of SYM perturbation 
theory \cite{BMN,gross,zanon}, and at the
nonplanar genus-one level \cite{KPSS,Constable1,BKPSS,Constable2} also incorporating
important effects due to mixing of the planar BMN operators.
Further investigations of the BMN sector in SYM were carried out in 
\cite{arut,bianchi,n-point,parnachev,gursoy,beisert,minahan,b2}. 

The {\it mass--dimension} type duality relation was  
clarified and extended
in \cite{ver,gross2,bits2} 
(see also \cite{zhou,gomis})
where it was expressed in the form
\be \label{ham}
{H}_{\rm string} = {H}_{\rm SYM} - J.
 \ee
Here ${H}_{\rm string}$ is the full string field 
theory Hamiltonian, and ${H}_{\rm SYM} - J = \Delta-J$ is 
the gauge theory Hamiltonian
(the conformal dimension) minus the R-charge.  The relation \eqref{ham} is expected
to be exact and hold to all orders in the two free parameters of the theory, 
$g_2$ and $\l'$.

However, one can argue that if the the relation \eqref{ham} was all that there
is in the pp-wave/SYM correspondence, this would not add much to our understanding
of neither the interactions of the massive modes in string theory, nor to the
gauge theory dynamics in the large $N$ double scaling limit. It would also be
rather unsatisfactory aesthetically. Recall that 
in the original AdS/CFT proposal, in addition to the relation
between the masses of supergravity states (and their KK towers) and the conformal
dimensions of the dual operators in SYM, 
one could also compare directly the
correlation functions in gauge theory with the bulk interaction
vertices \cite{witten,gkp} using the bulk-boundary propagators. 
Since the pp-wave/CFT correspondence can
be considered as a particular limit of the AdS/CFT correspondence, 
it is natural to suspect that similar {\it vertex--correlator} 
type duality relation will hold in the pp-wave/SYM correspondence.
Since the attempts so far \cite{hr}
of uncovering similar to the AdS/CFT holographic relation
turned out to be extremely difficult and somewhat unfruitful, there is a need
for a different route to establish a dynamical 
vertex--correlator type pp-wave duality (if any).
In this paper we will use the large body of recently derived detailed field theory 
results for BMN correlation functions as the `experimental data' for building up
a theoretical model of such a relation.

An important observation that the 3-point
function in a conformal field theory takes a universal form was put
to use in \cite{Constable1,CKT}. 
In conformal theory, the two- and three-point functions of conformal primary 
operators are completely determined by conformal invariance of the
theory. One can always
choose a basis of primary operators such that the two-point functions
take the canonical form:
\begin{equation} \label{2pt}
\langle {\cO}_I (x_1) \bar\cO_J(x_2) \rangle = \frac{\d_{IJ}}{(4 \pi^2
x_{12}^2)^{\Delta_I}},
\end{equation}
and all the nontrivial information of the
three-point function is contained in the $x$-independent 
coefficient $C_{1 2 3}$:
\begin{equation} \label{3pt}
\langle \cO_{1}(x_1) \cO_{2}(x_2) \bar\cO_{3}(x_3) \rangle  =
\frac{C_{123} }
{(4\pi^2 x_{12}^2)^{\frac{\D_1+\D_2 -\D_3}{2}}
 (4\pi^2 x_{13}^2)^{\frac{\D_1+\D_3 -\D_2}{2}}
 (4\pi^2 x_{23}^2)^{\frac{\D_2+\D_3 -\D_1}{2}}},
\end{equation}
where $x_{12}^2: = (x_1-x_2)^2$.
Since the form of the $x$-dependence of conformal 3-point functions is universal,
one can identify the coefficient $C_{123}$ with a `coupling constant' of the
three BMN states in SYM.
It is then natural to expect that $C_{123}$
is related to the interaction of the corresponding three string states
in  the pp-wave
background \footnote{There is also a more technical reason for this relation:
we will show in Section 3 that all the available SYM results for
$\mu(\Delta_1+\Delta_2-\Delta_3)\, C_{123}$ can
be expressed  entirely in terms of the natural pp-wave string theory
quantities, such as Neumann matrices, oscillation frequencies etc.}.
Using the operator product expansion, this  3-point relation will then serve 
as the building block
for a string interpretation of $n$-point BMN correlators \cite{n-point}
in short distance limits.

A few general remarks are in order:

{\sl 1.}
We note that it is an essential part of our proposal to use on the SYM side 
the BMN operators defined in such a way that they do not mix 
with each other (i.e. have definite scaling
dimensions $\Delta$) and which are conformal primary operators. The BMN operators
defined in this way will be called the $\Delta$-BMN operators. Conformal invariance
of the $\cN=4$ theory then implies that the 2-point correlators 
of these $\Delta$-BMN 
operators are canonically normalized, 
and the 3-point functions take the simple form \eqref{3pt}. 

{\sl 2.} 
The relation \eq{ham} can be understood as the equivalence of the spectra
of the operators or in a
stronger form, as an operator equation. To
establish the latter, one would have to first establish an
isomorphism of the field theory and string theory Hilbert spaces, and
then compare the matrix elements of the operators in \eqref{ham}. This
point of view was adopted in, e.g. \cite{gross2,bits2}, 
where certain modified
BMN operator bases were considered.
Each of the two bases of \cite{gross2} and \cite{bits2} was reported to be
isomorphic to the Hilbert space of {\it bare string states}, i.e. the basis which
one uses to write down the tree-level 3-string vertex.
By construction, the bases of
\cite{gross2,bits2} were not the eigenstates of $\Delta$, and hence
different from the $\Delta$-BMN basis which we use here. 
Because of this, each of the bases of \cite{gross2,bits2}
was made orthonormal only at the {\it free} field theory level.
However at the interacting level ($\l'\neq0$) the 2-point functions of
the operators in \cite{gross2,bits2} will contain 
a non-universal logarithmic coordinate dependence.
It is not clear 
to us how to remove this dependence from the 2-point functions
and to define a coordinate-independent overlaps, unless one is using the
$\Delta$-BMN basis, where the coordinate dependence is universal, i.e. dictated
by \eqref{2pt}.

{\sl 3.} 
In this paper we are not attempting to construct the isomorphism between
the states in string theory and in the BMN sector of SYM.
For example, our $\Delta$-BMN basis\footnote{It can be used, however,
and is well-suited   
for calculating the spectum of \eq{ham} on the SYM side.}
is not isomorphic to the natural basis of bare string states.
Our proposal is, instead, to relate the
naturally defined in conformal field theory coefficient $C_{123}$ with the tree
level string interaction of bare string states. For this purpose,
the $\Delta$-BMN basis is unique, as only for such a basis 
one can write down \eqref{3pt} and
the coefficient $C_{123}$ is defined.

The paper is organized as follows.
In Section 2, we will examine and elaborate on a specific vertex--correlator duality
relation (Eq. \eq{hh} below) originally proposed in \cite{Constable1}.  In
Section 3, we will subject this proposal to (twelve) tests. 
We will show that all the results for  field theory 3-point
functions that are available in the literature up to date,
including BMN operators with 2 impurities, \cite{CKT,BKPSS}
and with 3 impurities, \cite{GK},
can be precisely reproduced on the string theory side with a
specific choice of the string theory prefactor
${\sf P}$ on the right hand side of \eqref{hhhh}. We emphasize that this
matching is nontrivial even though within our approach the choice of the
prefactor is ``phenomenological''. A first principles
derivation of the string field theory prefactor is highly
desirable, but not yet available, \cite{z2-1,z2-2},
inspite of a recent progress \cite{HSSV} and much work
on the construction of the 3-string vertices
in the pp-wave lightcone string field theory
\cite{SV1,SV2,rel3,ari1,ari2,z2-1,z2-2,HSSV}.

In this paper we are not concerned with
fermionic BMN operators in gauge theory, hence we are not probing the fermionic structure of the 3-string vertex. Our prefactor is an effective bosonic part
of the full prefactor, it
is clearly $Z_2$ invariant, but we cannot study the full 
(super)-symmetry of the vertex
in the pp-wave background  without including fermions.

\section{The correspondence between field theory 3-point function and
3-string vertex}

The idea is to compare directly the
3-strings interaction amplitude with the field theory structure
coefficients via \cite{Constable1}
\be \label{hh}
 \mu(\Delta_1+\Delta_2-\Delta_3)\, C_{123} 
= \langle \Phi_1| \langle \Phi_2|\langle \Phi_3| H_{3} \rangle  .
\ee 
Here $\langle \Phi_1| \langle \Phi_2|\langle \Phi_3| H_{3} \rangle $ is
the three-string scattering amplitude 
in the string field theory
% $ \langle 0_1| \langle 0_2|\langle 0_3| H_{3}\rangle$ is the vacuum amplitude 
and  $\ket{H_{3}}$ is the lightcone three-string vertex. 
$C_{123}$ is
the three-point function coefficient in \eq{3pt} of the
corresponding BMN operators.
We propose that equation \eq{hh}
is valid to all orders in perturbation theory in the effective gauge coupling $\l'$
of the BMN sector,
\be \label{lampr}
 \l' = \frac{g_{\rm YM}^2 N}{J^2} = \frac{1}{(\mu p^+ \a')^2}\ 
\ee
and at the leading order in the field theory genus counting parameter
\begin{equation} \label{gtwo}
g_2 :=\frac{J^2}{N}= 4 \pi g_s (\mu p^+ \a')^2 \ .
\end{equation}

To proceed, we need to specify the expression for the 3-string vertex $\ket{H_3}$.
The 3-string vertex can be represented as a ket-state in the tensor product of
the three string  Fock spaces. It has the form
\be
\ket{H_3} = {\sf P}  \ket{V_F} \ket{V_B}\d(\sum_{\rr=1}^{3} \a_\rr),
\ee
where the kets $\ket{V_B}$ and  $\ket{V_F}$ are constructed to satisfy
the bosonic and fermionic kinematic symmetries and $\a_\rr$ are defined 
in \eqref{frame} in the Appendix.
The bosonic factor $\ket{V_B}$ is given by
\be \label{Vb}
\ket{V_B} = \exp( \frac{1}{2} \sum_{\rr,\rs=1}^3 \sum_{m,n=-\infty}^\infty
\sum_{I=1}^8 \a^{\rr\, I\dag}_{m} \Nh_{mn}^{\rr\rs}  \a^{\rs\,
I\dag}_{n}) \ket{0}_{123},
\ee
where the $\Nh_{mn}^{\rr\rs}$ are the Neumann matrices in the
BMN-basis of string oscillators (as defined in eqn. \eq{bmn-a}). 
To simplify our notation in what follows we suppress the explicit 
sum over the $I$ indices. 
The complete perturbative expansion of the Neumann matrices 
in the pp-wave background
in the vicinity of  $\mu=\infty$, 
was recently constructed in \cite{HSSV}
\footnote{We refer the reader to the Appendix
for some useful properties of the perturbative Neumann matrices, relations
between different string-oscillator bases, and
the comparison with other results in the literature.}.
The prefactor
${\sf P} $ is a polynomial in the bosonic and fermionic
oscillators and should be determined from imposing the remaining symmetries 
of the pp-wave background. The fermionic factor $\ket{V_F}$ 
is not going to be relevant for the present
paper where only external {\it bosonic} string states are considered. 

The construction of the 3-string vertex has been
considered in \cite{SV1,SV2,ari1,ari2}, however
as emphasized in \cite{z2-1,z2-2}, 
string theory in the pp-wave background must respect the full symmetry 
of the background,
including the bosonic
symmetry $SO(4)\times SO(4) \times Z_2$, where the $Z_2$ exchanges the
first $SO(4)$ with the second $SO(4)$. 
It turns out that the string interactions
constructed in \cite{SV1,SV2,ari1,ari2} 
do not respect the $Z_2$ symmetry and so
cannot fully describe the string interaction in the
pp-wave background. Implementation of the $Z_2$ symmetry at the level of the
fermionic overlap $\ket{V_F}$ has been performed in \cite{z2-1,z2-2}.
Explicit expression of $\ket{V_F}$ 
is given in eqn.(16)
of \cite{z2-2}. Based on this starting point, one can at least in principle
construct the  prefactor  ${\sf P}$ by imposing on the vertex dynamical 
(super)symmetries of the 
background. However the algebra is quite involved \cite{wp} and the explicit form 
for the prefactor has not yet
been determined from the first principles. 

In this paper we take a different approach and instead of deriving the prefactor
in string theory we propose a simple ansatz for the bosonic part of the
prefactor ${\sf P}$ which is then subjected
to numerous independent tests against
all the available field theory expressions.
Our  ansatz for the relevant bosonic part of the prefactor is 
\be \label{pf}
{\sf P} = C_{\rm norm}({\sf P}_{I}+{\sf P}_{II}) ,
\ee
where\footnote{Throughout the paper 
we are using the usual definitions for the SFT quantities
in the pp-wave background such as 
$\omega_{\rr m}$, $\alpha_\rr$ and $\mu$, which are summarized in the Appendix.}
\be \label{pf1}
{\sf P}_{I} = 
\sum_{\rr=1}^3\sum_{m=-\infty}^{+\infty}\frac{\omega_{\rr m}}{\alpha_\rr}\, 
\a^{\rr \, I\dag}_m \a^{\rr\, I}_m,
\ee
and
\be \label{pf2}
{\sf P}_{II} = \frac{1}{2}
\sum_{\rr,\rs=1}^{3}\sum_{m,n > 0}
\frac{\omega_{\rr m}}{\alpha_\rr}\,
(\hat{N}^{\rr\rs}_{m -n} - \hat{N}^{\rr\rs}_{m n})
(\a^{\rr\, I \dag}_m \a^{\rs\, I\dag}_n + 
\a^{\rr\, I \dag}_{-m} \a^{\rs\, I \dag}_{-n} - 
\a^{\rr\, I \dag}_{m} \a^{\rs\, I \dag}_{-n} - \a^{\rr\, I \dag}_{-m} 
\a^{\rs\, I\dag}_{n} )
\ee
Apart from an overall numerical coefficient,\footnote{Which turns out 
to be precisely equal to the structure constant
$C^{\rm vac}_{123}$
of the ``vacuum'' BMN operators 
$\cO^J_{\rm vac} := (J N^{J})^{-1/2}\, \tr (Z^J)$.} 
\be \label{cnorm}
C_{\rm norm} = g_2 \frac{\sqrt{y(1-y)}}{\sqrt{J}}= C^{\rm vac}_{123},
\ee
our prefactor \eqref{pf} 
is the sum of two contributions, ${\sf P}_{I}$ and ${\sf P}_{II}$.

The first contribution ${\sf P}_{I}$, given by \eqref{pf1}, 
is simply the difference between the light-cone energies
of the string states\footnote{In this paper we always assume that the
outgoing string state and one of the incoming string states are excited
states, i.e. have non-zero eigenvalues of the number operator 
$\a^{\dag}_m \a_m $.}
 or, equivalently, $\mu$ times the difference of 
the scaling dimensions of the
incoming and outgoing BMN operators, $\Delta_1+\Delta_2-\Delta_3$.
The second contribution ${\sf P}_{II}$, given in \eqref{pf2}, 
is also bilinear  in string oscillators,
but involves creation operators only. For future reference we also note that
the sum \eqref{pf2} 
excludes the supergravity modes $m,n=0$.

It is worthwhile to note that our ansatz for the prefactor, 
\eq{pf1} and \eqref{pf2}, takes a remarkably simple form when
expressed in terms of the original SFT $a$-oscillator basis
\footnote{To derive \eq{pf22}, we have used 
% \be \label{pf23}
$ {\sf P}_{II} = -
\sum_{\rr,\rs=1}^{3}\sum_{m,n > 0}
% \omega_{\rr m}/\alpha_\rr
\frac{\omega_{\rr m}}{\alpha_\rr}\,
{N}^{\rr\rs}_{-m -n} 
\,\, a^{\rr\, I \dag}_{-m} a^{\rs\, I\dag}_{-n},$
% \ee
which follows from \eq{pf2} directly,
the properties \eq{N-exact-prop2} and the fact that 
when acting on $\ket{V_B}$, $a$ acts like
a derivative with respect to $a^\dag$.}
\bea 
&& {\sf P}_{I} = \sum_{\rr =1}^{3} \left(
\sum_{m > 0} 
\frac{\omega_{\rr m}}{\alpha_\rr}\,
(a^{\rr\, I \dag}_{m} a^{\rr\, I}_{m}+ a^{\rr\, I \dag}_{-m} a^{\rr\, I}_{-m})
+ \mu \, \mbox{sign} (\a_\rr) a_0^{\rr\, I \dag} a_0^{\rr\, I}
\right), \label{pf11} 
\\
&& {\sf P}_{II} = -\sum_{\rr =1}^{3}\sum_{m > 0} \frac{\o_{\rr
    m}}{\a_\rr} a_{-m }^{\rr\, I\dag}  a_{-m }^{\rr\, I}.  
\label{pf22}
\eea
Hence the full prefactor is
\be
{\sf P} = \sum_{\rr =1}^{3} \left(
\sum_{m > 0} 
\frac{\omega_{\rr m}}{\alpha_\rr}\,
a^{\rr\, I \dag}_{m} a^{\rr\, I}_{m}
+ \mu \, \mbox{sign} (\a_\rr) a_0^{\rr\, I \dag} a_0^{\rr\, I}
\right).
\ee
This is different from the earlier proposals for the prefactor 
in \cite{SV2,ari1,HSSV}.
Although the prefactor takes a simpler form in the SFT $a$-oscillator
basis, we will continue using the prefactor 
in the BMN $\a$-oscillator basis, \eq{pf1} and \eq{pf2}, where the
comparison with the gauge theory BMN correlators is more direct. 

Our prefactor, and in particular the second term ${\sf P}_{II}$, is constructed to 
reproduce a particular class of field theory results\footnote{Namely
the two expressions considered in subsections 
3.3.1 and 3.3.2.} 
for the 3-point functions. 
It will turn out that the relatively simple expressions for ${\sf P}_{I}$ 
and ${\sf P}_{II}$ will match with {\it all} of the available field 
theory results, thus passing numerous
non-trivial tests detailed in Section 3. 
We also find it encouraging that 
the coefficient matrix in front of the oscillator-bilinear in \eqref{pf22} or
\eqref{pf2} is assembled directly from the Neumann matrices
rather than being given by a generic function of
$m,n$ and $\rr,\rs$. Since the Neumann matrices are known \cite{HSSV}
to all orders in the
perturbative expansion in inverse powers of $\mu$, our proposed
correspondence provides
an all-orders in $\lambda'$ prediction for the 3-point BMN correlators in SYM.

To summarize: for the bosonic external string states $\langle \Phi_i|$
our proposed correspondence relation is
\be \label{hhhh}
\mu(\Delta_1+\Delta_2-\Delta_3)\, C_{123}
= 
\langle \Phi_1| \langle \Phi_2|\langle \Phi_3| \, {\sf P}\, 
\exp( \frac{1}{2} \sum_{\rr,\rs=1}^3 \sum_{m,n=-\infty}^\infty
\sum_{I=1}^8 \a^{\rr\, I\dag}_{m} \Nh_{mn}^{\rr\rs}  
\a^{\rs\, I\dag}_{n}) \,\ket{0}_{123},
\ee 
where ${\sf P}$ is given by \eqref{pf} and $\Nh_{mn}$ are the perturbative 
Neumann matrices of \cite{HSSV} detailed in the Appendix.

\section{Tests of the Proposal}

As explained earlier, on the SYM side of our proposed correspondence 
we must use the $\Delta$-BMN operator basis. For BMN operators with 
2 scalar impurities this basis was
constructed in \cite{BKPSS} to order ${g_2} (\l')^0$ and
${g_2}^2(\l')^0$ and involves a linear combination
of the original single-trace BMN operator and the double-trace 
(in general multi-trace)
BMN operators.

\subsection{SYM predictions}

All the currently known SYM results for 3-point BMN correlators can be 
divided into two
broad classes. The {\it first class} involves 1 general and 2 chiral 
$\Delta$-BMN operators 
(i.e. 1 string and 2 supergravity states). Furthermore, 
no flavour-changing processes
are allowed for the 3-point functions of the first class.
The {\it second class} involves 2 general
$\Delta$-BMN operators (i.e. 2 string states and 
1 supergravity state) with or without
flavour changing, and the flavour changing 1 string $\to$ 2 supergravity processes.

The 3-point functions of the first class can be calculated directly 
with the original
single-trace BMN operators since it is easy to check that the contributions 
coming from the
double-trace operators vanish (at the first non-vanishing order in $g_2$). 
Three-point functions
of the first class were calculated in \cite{CKT} for 2 scalar impurities and, in
the follow-up paper
\cite{GK}, for 3 scalar impurities at the leading non-trivial order in $g_2$.
For the second class the contributions from the double-trace operators are 
important and one should use the $\Delta$-BMN basis. The calculations 
of the 3-point functions 
in this basis with 2 scalar impurities were carried out in \cite{BKPSS}.

All the currently known
SYM results for 3-point correlators with 2 scalar impurities can be 
summarized by
the following expression for the 3-point function coefficients $C_{123}$:
\be \label{twoi1}
C( k_{n}l_{-n},\, {\rm vac}|\, i_m j_{-m})  =
C_{123}^{\rm vac}\frac{2\,\sin^2(\pi m y)}{y\, \pi^2 (m^2-n^2/y^2)^2}
\lrbrk{\delta_{i(k}\delta_{l)j}\,\,m^2+\delta_{i[k}\delta_{l]j}\,\frac{m n}{y}+
\sfrac{1}{4}\delta_{ij}\delta_{kl}\, \frac{n^2}{y^2}} , 
\ee
\begin{equation} \label{twoi2}
C(k_0,\,l_0 |\, i_m j_{-m}) 
 = C_{123}^{\rm vac}
\frac{2}{\sqrt{y(1-y)}}\lrbrk{\delta_{m,0}\, y
-\frac{\sin^2(\pi m y)}{\pi^2 m^2}} \delta_{i(k}\delta_{l)j},
\end{equation}
where $C(1,2|3)$ is the coefficient for the 3-point function 
$\langle  \cO_1^{J_1} \cO_2^{J_2} \bar{\cO}_3^{J}\rangle$, $J= J_1 +J_2$ and
\be
y:= J_1/J
\ee 
is the R-charge ratio.
Here the 
``string modes'' $m$ and $n$ can be positive, negative or zero, 
impurities flavours
$i,j,k$ and $l$ are arbitrary integers from the set $\{1,2,3,4\}$, 
and the symmetric traceless and the antisymmetric traceless combination of the
two Kroneckers are defined as
\begin{equation}
\delta_{i(k}\delta_{l)j}= \sfrac{1}{2}(\delta_{ik} 
\delta_{lj} +\delta_{il} \delta_{kj}) 
- \sfrac{1}{4}\delta_{ij}\delta_{kl} \ , \quad
\delta_{i[k}\delta_{l]j}= \sfrac{1}{2}(\delta_{ik} 
\delta_{lj} -\delta_{il} \delta_{kj}) .
\label{sdel}
\end{equation}
When $n=0$ and $i,j=k,l$ or $i,j=l,k$,
these 3-point functions are from the first class and were originally
derived in \cite{CKT}.
Otherwise these results are from the second class and were derived
in \cite{BKPSS}.
It is important to note that the calculations of \cite{CKT,BKPSS} were
performed to order $(\l')^1$, incorporating the leading order in
$\l'$ anomalous dimensions of the $\Delta$-BMN operators in \eqref{3pt}.
However, the resulting expressions for the three-point 
coefficients $C_{123}$ can be
trusted only to order $(\l')^0$ \cite{BKPSS}. This is because the yet unaccounted
order $g_2 (\l')^1$ corrections to the mixing 
matrices of the single- and the double-trace
operators will affect the order $(\l')^1$ expressions for $C_{123}$ (but not the
logarithmic anomalous dimensions). Hence, the expressions on the right hand side of
\eqref{twoi1}, \eqref{twoi2} are 
given at order $(\l')^0$. We stress that they are different from the naive free
field theory results as they already incorporate the operator mixing at order
$g_2 (\l')^0$.

To go beyond these SYM results involving the 3-point $\Delta$-BMN functions with
2 scalar impurities one can consider, for example, BMN operators with 3 scalar
impurities $i,j,k$ with string oscillator numbers $n_i,n_j,n_k$ satisfying the
constraint $n_i+n_j+n_k =0$,
\begin{eqnarray}
&&\cO_{i_{n_i} j_{n_{j}}k_{n_{k}}} = \\
&&\frac{1}{J \sqrt{N^{J+3}}} \sum_{0\le a,b }^{a+b\le J}
\left[\tr \left( Z^a \phi_j Z^b \phi_k Z^{J-a-b}\phi_i \right)\, 
q_j^a q_k^{a+b}
+\tr \left( Z^a \phi_k Z^b \phi_j Z^{J-a-b} \phi_i\right)\, 
q_k^a q_j^{a+b} \right],  \nn
\end{eqnarray}
where $q_j=e^{2\pi i n_j/J}$ and $q_k=e^{2\pi i n_k/J}$ are the BMN phase-factors.
Three-point functions involving of $\Delta$-BMN operators with 3 scalar impurities
were evaluated very recently and will be reported in detail 
in the follow-up paper \cite{GK}.
A simple example of a more general construction
in \cite{GK} involves a 3-point function of
1 string-state BMN operator 
$\bar\cO_{i_{n_i}j_{n_{j}}k_{n_{k}}}$, and 2 supergravity-state operators,
$\cO_{i_0,j_0,k_0}$ and $\cO_{\rm vac}$, which is of conformal form \eqref{3pt},
giving the result for the coefficient:
\be 
C(1_0 2_0 3_0, \, {\rm vac} |\, 1_{n_1} 2_{n_2} 3_{n_3})
= 
- C_{123}^{\rm vac}\,
\frac{\sin(\pi y n_1)\sin(\pi y n_2)\sin(\pi yn_3)}
{y^{3/2} \pi^3 \, n_1 n_2 n_3} .
\label{threei}
\ee
Another simple example is a 3-point function of the same
1 string-state BMN operator 
$\bar\cO_{i_{n_i}j_{n_{j}}k_{n_{k}}}$, 
and 2 lower-impurity supergravity-state operators,
$\cO_{i_0,j_0}$ and $\cO_{k_0}$,
\be 
C( 1_0 2_0, \, 3_0 |\, 1_{n_1} 2_{n_2} 3_{n_3})
=  C_{123}^{\rm vac}\,
\frac{\sin(\pi y n_1)\sin(\pi y n_2)\sin(\pi yn_3)}
{y\sqrt{1-y} \pi^3 \, n_1 n_2 n_3} .
\label{threeitwo}
\ee

Having collected the field theory results above, we are now
ready to perform explicit tests of our proposed correspondence.
In doing so we will be testing the duality relation \eq{hhhh} itself, 
our ansatz for the prefactor \eqref{pf},\eqref{pf1},\eqref{pf2},\eqref{cnorm} 
and the expressions for the Neumann matrices \cite{HSSV} from the Appendix. 

We will now split the corresponding string interactions according to the
quantum numbers of the external states. From now on we will always assume
that the different impurity indices $i$ and $j$ always take 
different values, $i\neq j$ 
and the same impurities will be denoted explicitly as $i$ and $i$. Also the
oscillator modes $m$ and $n$ will always be positive, the negative modes will be
denoted as $-m$ and $-n$, and the supergravity states as $0$.

\subsection{Two supergravity states and one string state with two impurities}

On the SYM side the `supergravity' mode $n=0$ and at large $\mu$ we have 
$\mu(\Delta_1+\Delta_2-\Delta_3)=-m^2/\mu$. On the SFT side
for two supergravity and one string state process ${\sf P}_{II}$ cannot
contribute since it includes the contributions only from non-zero modes.
Hence, only ${\sf P}_I$ contributes which is a 
diagonal operator with the eigenvalues
\be
{\sf P}_I=\mu(\Delta_1+\Delta_2-\Delta_3)=-\frac{m^2}{\mu}. 
\ee
 
\subsubsection{$\bf \quad i_0\, j_0\, +\, vac\, \to \, i_m\, j_{-m} $}

Here there are 2 supergravity states with no flavour-changing, 
hence the process belongs to the first class.h
The SYM prediction \eqref{twoi1} is
\be \label{symone}
\mu(\Delta_1+\Delta_2-\Delta_3) C_{123} =\, 
-C_{123}^{\rm vac}\, 
\frac{\sin^2(\pi my)}{y\, \pi^2 \mu} .
\ee
The product of the external string states is
\be
{}_{123}\langle 0 | \a^{3i}_m \a^{3j}_{-m} \a^{1i}_0 \a^{1j}_{0}, 
\ee
and the relevant part of the exponent in \eqref{Vuse} reads
\be
\exp [ \hat{N}_{m0}^{31}(\a^{3 i\dag}_{m}\a^{1i\dag}_{0}+
\a^{3j\dag}_{-m}\a^{1j\dag}_{0})].
\label{over2}
\ee
The resulting string field theory expression, cf \eqref{hhhh}, is
\be  
-\,\frac{m^2}{\mu}\, C_{123}^{\rm vac}\, \hat{N}_{m0}^{31}\, \hat{N}_{m0}^{31}=
-\,\frac{m^2}{\mu}\,C_{123}^{\rm vac}\, \frac{\sin^2(\pi m y)}{\pi^2 m^2 y},
\ee
which reproduces the gauge theory result \eqref{symone}.

\subsubsection{$\bf \quad  i_0\, +\, j_0\, \to \, i_m\, j_{-m} $}

This is also the first class process.
The gauge theory expression \eqref{twoi2} gives
\be \label{symtwo}
\mu(\Delta_1+\Delta_2-\Delta_3) C_{123} =\, 
C_{123}^{\rm vac}\, 
\frac{\sin^2(\pi my)}{\sqrt{y(1-y)}\, \pi^2\mu} .
\ee
The product of the external string states is now
\be
{}_{123}\langle 0 |\a^{3i}_m \a^{3j}_{-m} \a^{1i}_0 \a^{2j}_{0}, 
\ee
and the relevant part of the exponent is again \eqref{over2}.
This gives the string prediction
\be  
-\,\frac{m^2}{\mu}\,C_{123}^{\rm vac}\, \hat{N}_{m0}^{32}\, \hat{N}_{m0}^{31}=
-\,\frac{m^2}{\mu}\,C_{123}^{\rm vac}\, 
\left(\frac{\sin(\pi m y)}{\pi m \sqrt{1-y}}\right)
\left(-\,\frac{\sin(\pi m y)}{\pi m \sqrt{y}}\right),
\ee
which is in agreement with the SYM expression \eqref{symtwo}.

\subsubsection{$\bf \quad j_0\, j_0\, +\, vac\, \to \, i_m\, i_{-m} $}

This is a flavour changing process since $i\neq j$ and it belongs 
to the second class.
The SYM prediction \eqref{twoi1} is
\be \label{symthree}
\mu(\Delta_1+\Delta_2-\Delta_3) C_{123} =\, 
\frac{1}{2}\, C_{123}^{\rm vac}\, 
\frac{\sin^2(\pi my)}{y\, \pi^2\mu} .
\ee
The product of the external string states is
\be
{}_{123}\langle 0 |\a^{3i}_m \a^{3i}_{-m} \a^{1j}_0 \a^{1j}_{0}.
\ee
Since this is a flavour changing process, the leading order in $1/\mu$ contribution
comes from 
\be \label{pf2fch}
{\sf P}_{II} = 
-\frac{\omega_{3 m}}{\alpha_3}\,
(\hat{N}^{33}_{m -m} - \hat{N}^{33}_{m m})
(\a^{3i \dag}_m \a^{3i\dag}_{-m} )=
\frac{2}{\pi} \sin^2(\pi my)
(\a^{3i \dag}_m \a^{3i\dag}_{-m} ),
\ee
while ${\sf P}_{I}$ does not contribute.
In deriving \eqref{pf2fch} we have used the fact that 
\be
\hat{N}^{33}_{m -m} - \hat{N}^{33}_{m m}= 
N^{33}_{-m -m}
= \frac{2}{\mu \pi} \sin^2(\pi my) \propto \frac{1}{\mu},
\ee
and 
\be
\hat{N}^{\rr \rr}_{m -m} - \hat{N}^{\rr \rr}_{m m}= N^{\rr \rr}_{-m -m} =
\cO\left(\frac{1}{\mu^3}\right) \, , \quad {\rm for} \quad \rr=1,2.
\ee From 
the exponent \eqref{Vuse} in \eqref{hhhh} we get the factor of
\be \label{pfkkfch}
\hat{N}^{11}_{00}\,  
\a^{1j \dag}_0 \a^{1j\dag}_{0} =
\frac{1}{\mu\, 4 \pi y} \a^{1j \dag}_0 \a^{1j\dag}_{0} .
\ee
Substituting these expressions into \eqref{hhhh} we get the string theory prediction
\be \label{twentykkk}
\frac{1}{2}\, C_{123}^{\rm vac}\, 
\frac{\sin^2(\pi my)}{y\, \pi^2\mu},
\ee
which is precisely the SYM result \eqref{symthree}.

\subsubsection{$\bf \quad i_0\, i_0\, +\, vac\, \to \, i_m\, i_{-m} $}

The SYM result \eqref{twoi1} is
\be \label{symfour}
\mu(\Delta_1+\Delta_2-\Delta_3) C_{123} =\, 
-\frac{3}{2}\, C_{123}^{\rm vac}\, 
\frac{\sin^2(\pi my)}{y\, \pi^2\mu} .
\ee
The string theory prediction is equal twice the contribution 
from the subsection 3.2.1
plus the contribution
from the subsection 3.2.3, 
which amounts precisely to
the right hand side of \eqref{symfour}.

\subsubsection{$\bf \quad j_0\, +\, j_0\, \to \, i_m\, i_{-m} $}

The SYM prediction \eqref{twoi2} is
\be \label{symfive}
\mu(\Delta_1+\Delta_2-\Delta_3) C_{123} =\, 
-\frac{1}{2}\, C_{123}^{\rm vac}\, 
\frac{\sin^2(\pi my)}{\sqrt{y(1-y)}\, \pi^2\mu} .
\ee
The product of the external string states is
\be
{}_{123}\langle 0 | \a^{3i}_m \a^{3i}_{-m} \a^{1j}_0 \a^{2j}_{0}.
\ee
This is a flavour changing process, ${\sf P}_{I}$ does not contribute and the 
leading order in $1/\mu$ contribution
comes from ${\sf P}_{II}$ given by \eqref{pf2fch}. From the 
exponent \eqref{Vuse} in \eqref{hhhh} we get the factor of
\be \label{pf4fch}
\hat{N}^{21}_{00}\,  
\a^{1j \dag}_0 \a^{1j\dag}_{0} =
-\frac{1}{\mu 4 \pi \sqrt{y(1-y)}} \a^{1j \dag}_0 \a^{1j\dag}_{0} .
\ee
Substituting these expressions into \eqref{hhhh} 
we get the string theory prediction
\be \label{twentyone}
-\frac{1}{2}\, C_{123}^{\rm vac}\, 
\frac{\sin^2(\pi my)}{\sqrt{y(1-y)}\, \pi^2\mu},
\ee
which is precisely the SYM result \eqref{symfive}.

\subsubsection{$\bf \quad i_0\, +\, i_0\, \to \, i_m\, i_{-m} $}

The SYM prediction \eqref{twoi2} is
\be \label{symsix}
\mu(\Delta_1+\Delta_2-\Delta_3) C_{123} =\, 
\frac{3}{2}\, C_{123}^{\rm vac}\, 
\frac{\sin^2(\pi my)}{\sqrt{y(1-y)}\, \pi^2\mu}.
\ee
It is easy to see that
the corresponding string prediction is equal to twice the string prediction
of \eqref{symtwo} plus the string prediction of \eqref{twentyone}.
This is in perfect agreement with the SYM result \eqref{symsix}.

\subsection{Two string states with two impurities}

On the SYM side at large $\mu$ we have 
$\mu(\Delta_1+\Delta_2-\Delta_3)=-(m^2-n^2/y^2)/\mu$. On the SFT side
this is matched by ${\sf P}_I=\mu(\Delta_1+\Delta_2-\Delta_3)=-(m^2-n^2/y^2)/\mu$,
but now also the second term in the prefactor, ${\sf P}_{II}$, has
to be taken into account.

\subsubsection{$\bf \quad i_{n}\, j_{-n}\, +\, vac\, \to\, i_m\, j_{-m} $}

The SYM prediction \eqref{twoi1} is
\be \label{symseven}
\mu(\Delta_1+\Delta_2-\Delta_3) C_{123} =\, 
-C_{123}^{\rm vac}\, 
\frac{\sin^2(\pi my)}{y\, \pi^2 \mu}\, 
\frac{m^2+\frac{mn}{y}}{m^2 -\frac{n^2}{y^2}}.
\ee
The external string states are ${}_{123}\langle 0 |
\a^{3i}_m \a^{3j}_{-m} \a^{1i}_n \a^{1j}_{-n}$
and the contributing terms in ${\sf P}_{II}$ take form, cf \eqref{pf2}
\be \label{pfseven}
{\sf P}_{II} = \frac{1}{2}
\left(\frac{\omega_{3 m}}{\alpha_3}+\frac{\omega_{1 n}}{\alpha_1}\right)\,
(\hat{N}^{31}_{m -n} - \hat{N}^{31}_{m n})
(\a^{3i \dag}_m \a^{1i\dag}_n + 
\a^{3j \dag}_{-m} \a^{1j \dag}_{-n} ).
\ee
At large $\mu$
\be \label{omreln}
\frac{\omega_{3 m}}{\alpha_3}+\frac{\omega_{1 n}}{\alpha_1}=\, 
-\frac{1}{2\mu}\, \left(m^2-\frac{n^2}{y^2}\right) +\, 
\cO\left(\frac{1}{\mu^3}\right),
\ee
we have 
\be 
{\sf P}_{II} = -\frac{1}{4\mu}\, \left(m^2-\frac{n^2}{y^2}\right)
\,
(\hat{N}^{31}_{m -n} - \hat{N}^{31}_{m n})
(\a^{3i \dag}_m \a^{1i\dag}_n + 
\a^{3j \dag}_{-m} \a^{1j \dag}_{-n} ).
\ee
The first term in the prefactor gives
\be 
{\sf P}_{I} = -\frac{1}{\mu}\, \left(m^2-\frac{n^2}{y^2}\right).
\ee
Combining these expressions together with $C_{123}^{\rm vac}$ in \eqref{pf} and
with the external states and the exponent \eqref{Vuse} we get
the string theory answer \eqref{hhhh}
\be 
 \frac{C_{123}^{\rm vac}}{\mu} \left(\frac{n^2}{y^2}-m^2\right)
\left[\frac{1}{2}(\hat{N}^{31}_{m -n} - \hat{N}^{31}_{m n}) + 
\hat{N}^{31}_{mn} 
\right] \hat{N}^{31}_{mn}=
\frac{C_{123}^{\rm vac}}{2\mu} \left(\frac{n^2}{y^2}-m^2\right)
(\hat{N}^{31}_{m -n} +\hat{N}^{31}_{m n})
\hat{N}^{31}_{mn} .
\ee  
Finally, using the expressions for the Neumann matrices from the Appendix
\be
(\hat{N}^{31}_{m -n} +\hat{N}^{31}_{m n})
\hat{N}^{31}_{mn}=
\frac{2}{\pi^2}
\frac{\sin^2(\pi m y)}{y(m^2-n^2/y^2)^2}\left(m^2+\frac{mn}{y}\right),
\ee
we derive the right hand side of the SYM expression \eqref{symseven}.

\subsubsection{$\bf \quad j_{n}\, i_{-n}\, +\, vac\, \to \, i_m\, j_{-m} $}

The SYM result \eqref{twoi1} is
\be \label{symeight}
\mu(\Delta_1+\Delta_2-\Delta_3) C_{123} =\, 
-C_{123}^{\rm vac}\, 
\frac{\sin^2(\pi my)}{y\, \pi^2 \mu}\, 
\frac{m^2-\frac{mn}{y}}{m^2 -\frac{n^2}{y^2}}.
\ee
The string theory prediction is obtained as in the previous subsection,
except that now ${\sf P}_{II}$ receives contributions from the other two
oscillator bilinears,
\be 
{\sf P}_{II} = -\frac{1}{2}
\left(\frac{\omega_{3 m}}{\alpha_3}+\frac{\omega_{1 n}}{\alpha_1}\right)\,
(\hat{N}^{31}_{m -n} - \hat{N}^{31}_{m n})
(\a^{3i \dag}_m \a^{1i\dag}_{-n} + 
\a^{3j \dag}_{-m} \a^{1j \dag}_{n} ).
\ee
The net result is
\be 
 \frac{C_{123}^{\rm vac}}{\mu} \left(\frac{n^2}{y^2}-m^2\right)
\left[-\frac{1}{2}(\hat{N}^{31}_{m -n} - \hat{N}^{31}_{m n}) + 
\hat{N}^{31}_{m-n}\right]\hat{N}^{31}_{m-n},
\ee 
which, using the expressions for the Neumann matrices at large $\mu$,
agrees precisely with the SYM expression \eqref{symeight}.

\subsubsection{$\bf \quad j_{n}\, j_{-n}\, +\, vac \, \to\, i_m\, i_{-m}$}

The SYM prediction \eqref{twoi1} is
\be \label{symnine}
\mu(\Delta_1+\Delta_2-\Delta_3) C_{123} =\, 
\frac{1}{2}\, C_{123}^{\rm vac}\, 
\frac{\sin^2(\pi my)}{y\, \pi^2 \mu}.
\ee
The external string states are 
$ {}_{123}\langle 0| 
 \a^{3i}_m \a^{3i}_{-m} \a^{1j}_n \a^{1j}_{-n}$
This is the flavour changing process and the leading order in $1/\mu$ contribution
comes from ${\sf P}_{II}$ and is given by \eqref{pf2fch}.
The exponent in \eqref{Vuse} gives the factor of
\be \label{pf3fch}
\hat{N}^{11}_{n -n} 
\a^{1j \dag}_n \a^{1j\dag}_{-n} =
\frac{1}{\mu 4 \pi y} \a^{1j \dag}_n \a^{1j\dag}_{-n} .
\ee
Putting this all together in \eqref{hhhh} we get the string theory prediction
\be
\frac{1}{2}\, C_{123}^{\rm vac}\, 
\frac{\sin^2(\pi my)}{y\, \pi^2 \mu},
\ee
which is precisely the SYM result \eqref{symnine}.

\subsubsection{$\bf \quad i_{n}\, i_{-n}\, +\, vac\, \to\, i_m\, i_{-m} $}

The SYM prediction \eqref{twoi1} is
\be \label{symten}
\mu(\Delta_1+\Delta_2-\Delta_3) C_{123} =\, 
-\frac{1}{2}\, C_{123}^{\rm vac}\, 
\frac{\sin^2(\pi my)}{y\, \pi^2 \mu}\, 
\frac{3m^2+\frac{n^2}{y^2}}{m^2 -\frac{n^2}{y^2}}.
\ee
It is easy to see that
the corresponding string prediction is simply the sum of the string predictions
of the three previous subsections. 
This is again in agreement with SYM since
the right hand side of \eqref{symten} is equal to the sum of the right hand sides
of equations \eqref{symseven},\eqref{symeight} and \eqref{symnine}.

\subsection{Three impurities}

Both of the processes considered below involve 2 supergravity states and
are of the first class, consequently 
only the first term in the prefactor, ${\sf P}_I$, gives a nontrivial contribution:
\be
{\sf P}_I=\mu(\Delta_1+\Delta_2-\Delta_3)=-\frac{n_1^2+n_2^2+n_3^2}{\mu},
\ee
where $n_1=-(n_2+n_3)$.

\subsubsection{ $\bf  \quad 1_0\, 2_0\, 3_0 \, +\, vac
\to\, 1_{n_1}\, 2_{n_2}\, 3_{n_3}$}

The external string state is:
\be
{}_{123}\langle 0 | \a^{3i_1}_{n_1}\a^{3i_2}_{n_2} \a^{3i_3}_{n_3}
\a^{1i_1}_{0}\a^{1i_2}_{0}\a^{1i_3}_{0}
\ee
and the relevant part of the exponent in \eqref{Vuse}
gives
\be
\hat{N}^{31}_{n_1 0} \, \hat{N}^{31}_{n_2 0} \, \hat{N}^{31}_{n_3 0}. 
\label{cubic}
\ee
The resulting string prediction is
\be 
-\mu(\Delta_1+\Delta_2-\Delta_3)\, C_{123}^{\rm vac}\,
\frac{\sin(\pi y n_1)\sin(\pi y n_2)\sin(\pi yn_3)}
{y^{3/2} \pi^3 \, n_1 n_2 n_3} .
\label{bthreei}
\ee
which reproduces \eqref{threei} precisely.

\subsubsection{ $\bf \quad 1_0 2_0\,+\, 3_0 
\, \to\, 1_{n_1}\, 2_{n_2}\, 3_{n_3}$}

Here
the external string state is: 
\be
{}_{123}\langle 0 | \a^{3i_1}_{n_1}\a^{3i_2}_{n_2} \a^{3i_3}_{n_3}
\a^{1i_1}_{0}\a^{1i_2}_{0}\a^{2i_3}_{0}
\ee
and the bosonic overlap, \eqref{Vuse}, gives
gives
\be
\hat{N}^{31}_{n_1 0} \, \hat{N}^{31}_{n_2 0} \, \hat{N}^{32}_{n_3 0}. 
\label{cubic1}
\ee
The resulting string prediction is
\be 
\mu(\Delta_1+\Delta_2-\Delta_3)\, C_{123}^{\rm vac}\,
\frac{\sin(\pi y n_1)\sin(\pi y n_2)\sin(\pi yn_3)}
{y\sqrt{1-y} \pi^3 \, n_1 n_2 n_3} .
\label{bthreeitwo}
\ee
which precisely reproduces \eqref{threeitwo}.

\section*{Acknowledgements} 

VVK would like to thank the Department of Physics of the National
Tsing Hua University and the National Center of Theoretical Science (NCTS) at
Hsinchu, where this work started, for their hospitality. He is also grateful to
Catherine and Nicholas for their support and understanding throughout.
We thank George Georgiou, Simon Ross, Rodolfo Russo, 
Gabriele Travaglini and 
Herman Verlinde for useful discussions. We acknowledge grants from
ESPRC, Nuffield foundation, PPRAC of UK and NSC and NCTS of  Taiwan

\section*{Appendix: Neumann Matrices in the large-$\mu^2$ Perturbation Theory}

We first specify the notation and conventions used in pp-wave string field theory.
The combination $\a'p^+$ for the r-th string is denoted $\a_\rr$
and $\sum_{\rr=1}^3 \a_\rr =0$. As is standard in
the literature, we will choose a frame in which $\a_3=-1$
\be \label{frame}
\a_\rr = \a'p^+_{(\rr)} \, : \qquad \a_3=-1, \qquad \a_1=y, \qquad \a_2=1-y.
\ee
In terms of the $U(1)$ R-charges 
of the BMN operators in the 
three-point function, $\langle \cO_1^{J_1} \cO_2^{J_2} \bar\cO_3^{J}
\rangle$, 
where $J=J_1+J_2$, we have
\be
y=\frac{J_1}{J}, \qquad 1-y=\frac{J_2}{J}, \qquad 0<y<1.
\ee
The effective SYM coupling constant \eqref{lampr} in the frame \eqref{frame}
takes a simple form
\be \label{lamprsim}
 \l' =  \frac{1}{(\mu p^+ \a')^2}\ \equiv \frac{1}{(\mu \a_3)^2}\ 
= \frac{1}{\mu^2}.
\ee
Here $\mu$ is the mass parameter which appears in the pp-wave metric, in
the chosen frame it is dimensionless\footnote{It is $p^{+}\mu$ which is
invariant under longitudinal boosts and is frame-independent.} and the
expansion in powers of $1/\mu^2$ is equivalent to the perturbative
expansion in $\l'$. Finally the frequencies are defined via,
\be
\omega_{\rr m}= \sqrt{m^2+(\mu\a_\rr)^2}.
\ee
 
The infinite-dimensional Neumann matrices, $N_{mn}^{\rr\rs}$ are usually specified
in the original $a$-oscillator basis of the string field theory. 
In this basis
the bosonic overlap factor $\ket{V_B}$ of the 3-strings vertex is given by
\be \label{Vbab}
\ket{V_B} = \exp( \frac{1}{2} \sum_{\rr,\rs=1}^3 \sum_{m,n=-\infty}^\infty
 a^{\rr\, I\dag}_{m} N_{mn}^{\rr\rs}  a^{\rs\, I\dag}_{n}) \ket{0}.
\ee
However, for the purposes of the pp-wave/SYM correspondence it is more convenient
to use another, the so-called BMN or $\a$-basis of string oscillators,
which is in direct correspondence with the BMN operators in gauge theory. The two
bases are related as follows:
\be \label{bmn-a}
\a_n = \frac{1}{\sqrt{2}}(a_{|n|} - i \, {\rm sign}(n) a_{-|n|}),
\quad \a_0 = a_0,
\ee
and satisfy the same oscillator algebra
\be
[\a_m, \a_n^\dag] = \d_{mn}.
\ee
In this basis, the bosonic overlap factor \eqref{Vbab} in the vertex reads
\be \label{Vbal}
\ket{V_B} = \exp( \frac{1}{2} \sum_{\rr,\rs=1}^3 \sum_{m,n=-\infty}^\infty
\a^{\rr\, I\dag}_{m} \hat{N}_{mn}^{\rr\rs}  \a^{\rs\, I\dag}_{n}),
\ee
where $\Nh$ are the Neumann matrices in the $\a$-basis and
are related to the $N$'s via (here $m,n>0$):
\bea 
&& \Nh^{\rr\rs}_{mn} = \Nh^{\rr\rs}_{-m-n} := 
\frac{1}{2}(N^{\rr\rs}_{mn} -N^{\rr\rs}_{-m-n}), \\
&& \Nh^{\rr\rs}_{m-n} = \Nh^{\rr\rs}_{-m n} := 
\frac{1}{2}(N^{\rr\rs}_{mn} + N^{\rr\rs}_{-m-n}), \\
&& \Nh^{\rr\rs}_{m0} =\Nh^{\rr\rs}_{-m0} := \frac{1}{\sqrt{2}} 
N^{\rr\rs}_{m0}=
\Nh^{\rr\rs}_{0m} =\Nh^{\rr\rs}_{0-m}, \label{poszer}\\
&&  \Nh^{\rr\rs}_{00}:=N^{\rr\rs}_{00}.
\eea
To derive these expressions we have equated \eqref{Vbab} and \eqref{Vbal}, and
used the known properties of the original perturbative Neumann matrices:
\bea 
&& N^{\rr\rs}_{MN}=N^{\rs\rr}_{NM} , \quad {\rm for\, all\,} -\infty<M,N<+\infty, 
\label{N-exact-prop1}\\
&& N^{\rr\rs }_{-m 0} =0, \quad 
 N^{\rr\rs}_{m -n}= 0, \quad \mbox{for $m,n>0$}.
 \label{N-exact-prop2}
\eea
Making use of \eqref{poszer} and \eqref{N-exact-prop1}, we can write
\eqref{Vbal} as
\bea \label{Vuse}
\ket{V_B} = \exp [ \frac{1}{2} \sum_{\rr,\rs=1}^3 \sum_{m,n=1}^\infty 
&&(\hat{N}_{00}^{\rr\rs}\a^{\rr\, I\dag}_{0}\a^{\rs\, I\dag}_{0}+
2\hat{N}_{m0}^{\rr\rs}(\a^{\rr\, I\dag}_{m}\a^{\rs\, I\dag}_{0}+
\a^{\rr\, I\dag}_{-m}\a^{\rs\, I\dag}_{0})   \\
&&+\hat{N}_{mn}^{\rr\rs}(\a^{\rr\, I\dag}_{m}\a^{\rs\, I\dag}_{n}+
\a^{\rr\, I\dag}_{-m}\a^{\rs\, I\dag}_{-n})
+\hat{N}_{m-n}^{\rr\rs}(\a^{\rr\, I\dag}_{m}\a^{\rs\, I\dag}_{-n}+
\a^{\rr\, I\dag}_{-m}\a^{\rs\, I\dag}_{n})
].  \nn
\eea

We now present the explicit expressions for the Neumann matrices 
in the original $a$-basis obtained by expanding the results
of \cite{HSSV} in powers of $1/\mu^2$. These expressions are needed
for calculations in Section 3.
\bea
&&N^{31}_{mn} = \frac{2 (-1)^{m+n+1}}{\pi}
\frac{m \sin(\pi m y)}{\sqrt{y}(m^2-n^2/y^2)}+\cO\left(
\frac{1}{\mu^2}\right), \\
&&N^{32}_{mn} = \frac{2 (-1)^{m}}{\pi}
\frac{m \sin(\pi m y)}{\sqrt{1-y}(m^2-n^2/(1-y)^2)}+\cO\left(
\frac{1}{\mu^2}\right), \\
&&N^{21}_{mn} = \frac{1}{\mu}\,\frac{ (-1)^{n+1}}{2\pi}
\frac{1}{\sqrt{y(1-y)}}+\cO\left(\frac{1}{\mu^3}\right), \\
&&N^{33}_{mn} = \cO\left(\frac{1}{\mu^3}\right), \\
&&N^{11}_{mn} = \frac{1}{\mu}\,\frac{ (-1)^{m+n}}{2\pi}
\frac{1}{y}+\cO\left(\frac{1}{\mu^3}\right), \\
&&N^{22}_{mn} = \frac{1}{\mu}\,\frac{ 1}{2\pi}
\frac{1}{1-y}+\cO\left(\frac{1}{\mu^3}\right).
\eea

\bea
&&N^{31}_{-m-n} = \frac{2 (-1)^{m+n}}{\pi}
\frac{n \sin(\pi m y)}{y^{3/2}(m^2-n^2/y^2)}+\cO\left(\frac{1}{\mu^2}\right), \\
&&N^{32}_{-m-n} = \frac{2 (-1)^{m+1}}{\pi}
\frac{n \sin(\pi m y)}{(1-y)^{3/2}(m^2-n^2/(1-y)^2)}+\cO\left(
\frac{1}{\mu^2}\right), \\
&&N^{21}_{-m-n} = \cO\left(\frac{1}{\mu^3}\right), \\
&&N^{33}_{-m-n} = \frac{1}{\mu}\,
\frac{2(-1)^{m+n}}{\pi} \sin(\pi m y)\sin(\pi n y)
+ \cO\left(\frac{1}{\mu^3}\right), \\
&&N^{11}_{-m-n} = \cO\left(\frac{1}{\mu^3}\right), \\
&&N^{22}_{-m-n} = \cO\left(\frac{1}{\mu^3}\right).
\eea

\bea
&&N^{33}_{00} = 0, \quad N^{31}_{00} =-\sqrt{y}, \quad 
N^{32}_{00} =-\sqrt{1-y}, \\
&&N^{12}_{00} = \frac{1}{\mu}\,\frac{(-1)}{4\pi}\frac{1}{\sqrt{y(1-y)}}  
=\, N^{21}_{00}, \\
&&N^{11}_{00} = \frac{1}{\mu}\,\frac{1}{4\pi}
\frac{1}{y}, \\
&&N^{22}_{00} = \frac{1}{\mu}\,\frac{1}{4\pi}
\frac{1}{1-y}.
\eea
For the zero-positive Neumann matrices 
%below we have divided
%the expressions presented in \cite{HSSV} by a factor of 2 in order
%to have the expressions below to be in agreement with the known expressions
%in the infinite $\mu$ limit.
% exprs below AGREE WITH HSSV version 2 
we have
\be
N^{31}_{0n}= 0, \quad N^{32 }_{0n} =0, \quad N^{33 }_{0n} =0.
\ee
\bea
&&N^{13}_{0n} = \frac{\sqrt{2} (-1)^{n+1}}{\pi}
\frac{\sin(\pi n y)}{n\sqrt{y}}+\cO\left(\frac{1}{\mu^2}\right), \\
&&N^{23}_{0n} = \frac{\sqrt{2} (-1)^{n}}{\pi}
\frac{\sin(\pi n y)}{n\sqrt{1-y}}+\cO\left(\frac{1}{\mu^2}\right), \\
&&N^{21}_{0n} = \frac{1}{\mu}\,\frac{\sqrt{2}(-1)^{n+1} }{4\pi}
\frac{1}{ \sqrt{y(1-y)}}+\cO\left(\frac{1}{\mu^3}\right), \\
&&N^{12}_{0n} = -\frac{1}{\mu}\,\frac{\sqrt{2}}{4\pi}
\frac{1}{\sqrt{y(1-y)}}+\cO\left(\frac{1}{\mu^3}\right), \\
&&N^{11}_{0n} = \frac{1}{\mu}\,\frac{\sqrt{2} (-1)^{n}}{4\pi}
\frac{1}{y}+\cO\left(\frac{1}{\mu^3}\right), \\
&&N^{22}_{0n} = \frac{1}{\mu}\,\frac{\sqrt{2}}{4\pi}
\frac{1}{1-y}+\cO\left(\frac{1}{\mu^3}\right).
\eea

The behavior of Neumann matrices at $\mu\to \infty$ was first analyzed in
\cite{CKT} by resumming all-orders power expansions in the large $\mu$ (small $\l'$)
limit. In manipulating with multiplications of infinite dimensional matrices 
the authors of \cite{CKT} encountered divergences which were regularized using
the zeta-function regularization. Recently in \cite{HSSV}
the Neumann matrices at $\mu\to \infty$
were calculated using a different method leading to manifestly regular expressions.
The results of \cite{HSSV} which we use in this paper
agree with the expressions obtained \cite{CKT} at order $(1/\mu)^0$, but 
not at higher orders in $1/\mu$.
Two comments are in order:

{\sl 1.}  These perturbative expressions for the Neumann matrices should not be 
interpolated to the flat space expressions at $\mu=0$ since essential
singularities at $\mu =\infty$ were discarded.

{\sl 2.} Each of the Neumann matrices is expanded in powers of $1/\mu^2$,
the odd powers of $\mu$ can appear only as an overall multiplicative factor.
Hence the fractional powers of $\l'$ hopefully should not appear in the
string theory prediction at higher orders, thus making a happy connection
with the gauge theory interpretation.

\end{document}